# Intelligent Human Machine Interface Design for Advanced Product Life Cycle Management Systems

Zeeshan Ahmed
Vienna University of Technology
Getreidemarkt 9/307, 1060 Vienna
Austria
Email: zeeshan.ahmed@tuwien.ac.at

## ABSTRACT
Designing and implementing an intelligent and user friendly human machine interface for any kind of software or hardware oriented application is always be a challenging task for the designers and developers because it is very difficult to understand the psychology of the user, nature of the work and best suit of the environment. This research paper is basically about to propose an intelligent, flexible and user friendly machine interface for Product Life Cycle Management products or PDM Systems since studies show that usability and human computer interaction issues are a major cause of acceptance problems introducing or using such systems. Going into details of the proposition, we present prototype implementations about theme based on design requirements, designed designs and technologies involved for the development of human machine interface.

## Categories and Subject Descriptors
[**Graphical User Interface**]:

## 1. INTRODUCTION
Product Data Management is way of managing all relevant data to improve the quality of products and followed processes. Product Data Management systems mainly manage information about design and manufacturing of products including technical operations and running projects. Though Product Data Management systems are heavily benefiting industry the Product Data Management community is also facing some serious unresolved issues .i.e., enterprise spanning Product Data Management system deployment, static and unfriendly user interfaces and unintelligent search mechanisms.

Targeting some of above mentioned issues .i.e., static and unfriendly graphical user interface, metadata based static and unintelligent search, we have proposed an approach called Intelligent Semantic Oriented Search (I-SOAS) [1] (See Figure 1).I-SOAS is an information engineering approach to provide a solution implementing intelligent application capable of handling user's structured and unstructured requests by processing, modeling and managing the data on a semantic level.

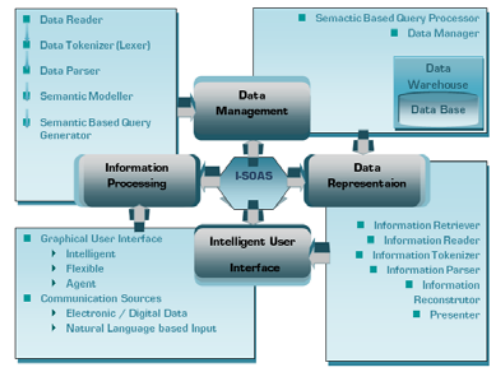

**Figure.1.** I-SOAS Conceptual Architecture [1]

To meet aforementioned goals I-SOAS's conceptual architecture is divided into four sequential iterative components .i.e., Intelligent User Interface, Information Processing, Data Management and Data Representation (See Figure 1). Intelligent User Interface is proposed to design intelligent human machine interface for system user communication, IP is proposed to process and model user's unstructured and structured inputted request by reading, lexing, parsing, and semantic modeling, Data Management is proposed to manage user request and system performance based information and Data Representation is proposed to represent system outputted results in user's understandable format [2].

In this research paper we are focusing on Intelligent User Interface. Moreover we also present information about involved tools and technologies in the development of I-SOAS Graphical.

## 2. DESIGN REQUIREMENTS
The I-SOAS Intelligent User Interface is designed to meet four major requirements .i.e., flexibility, agent based design, context awareness and natural language communication mode.

## 2.1 Flexibility

The designed graphical user interface must be flexible in order to enable users to redesign and reconfigure the interface itself to accommodate specific needs. Examples:

1. When software loads first time, then it should identify the level of user, whether it's a beginner or experienced. If the user is beginner than an automatic GUI trainer should be provided, which will explain the provided available options in the application, their usage and GUI design operations in best possible short time, and If the user is an experienced person, then will move user to the application directly.

2. User will be provided with two major options .i.e., Manual Option Adjustment and Standard GUI. Manual Option Adjustment will allow user to create a new GUI by providing several operations .i.e., Mouse Click and Drag Drop, Key-board Key Press, Screen Touch and Voice Recognition etc.

3. User will be provided with the option to adjust the available already defined GUI, save redesigned GUI as template, load or reload already designed template, delete running template and lock or unlock GUI.

4. User will be provided some options in the form of Standard GUI Controls like Buttons, Labels, List Box, Combo Box, Text Box etc., Color Schemes, Font, Graphics, Animations and Standard GUIs like forms to take data inputs and view the results etc.

5. User will be provided a desktop and web based platform independent graphical user interface.

## 2.2 Agent Based Design

The proposed graphical interface must contain a Personal Assistant, which should be capable to the following tasks .i.e.,

1. Personal Assistant should be able to move in the graphical interface, and this movement should be of two types .i.e., Explicit Movement (PA can be moved by the user) and Implicit Movement (PA can move itself according to the environmental adjustments).

2. Personal Assistant can be controlled and instructed by user using mouse click, drag drop, keyboard key press, screen touch and voice recognition options.

3. Personal Assistant should be able to produce emotions according to the nature of job and current situation like Typing on keyboard when user is typing or asking PA to type something for him, Listening when user is speaking and asking PA for some information or job, wear glasses when user is asking PA to look for something etc.

4. Personal Assistant should be able to save his state and maintain his history.

## 2.3 Context Awareness

The proposed graphical interface must perform the following jobs .i.e.,

1. Learn from experience
2. Help user in decision making
3. Provide Intelligent graphical user interface trainer to intelligently train the new user
4. Should be able to handle graphical user interface itself
5. Graphical user interface must be able to present output of inputted instruction in best possible way with respect to the user, like if user is asking in voice output should be delivered in voice etc.

## 2.4 Natural Communication Mode

The graphical interface must contain a natural communication mode, where user can instruct the software with natural language based instructions like user can write instruction in natural speaking language and software can read, understand the context and semantic, and then performed required action.

## 3. DRAFT - GRAPHICAL INTERFACE

The proposed graphical design, shown in Figure 2, is designed by following already mentioned designed requirements and it is based on Human Machine Interface Design concepts .i.e., Ideologies, Principles, Patterns and Guidelines [3].

This manual physical sketch is basically based on several different visible and non visible options but mainly there are two different categories of options .i.e., Control Interface and Personal Assistant.

### 3.1 Control Interface

The designed manual physical sketch of Graphical Interface mainly consists of a Control Container. Control Container is the main front page of the graphical user interface which contains all the control options including list boxes, mouse hover/click, drag drop, drop down list boxes, list boxes and Personal Assistant. These options are provided to the user to perform certain tasks. Moreover user can redesign or make changes in already designed

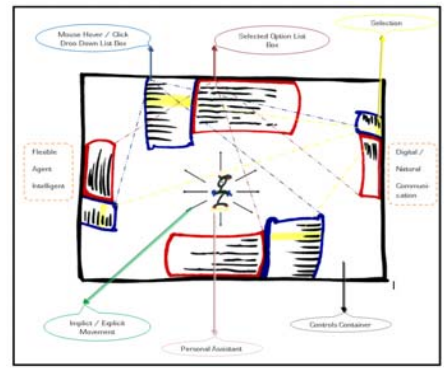

**Figure.2.** I-SOAS: Graphical Interface Draft Design

graphical interface using these options and controls (see Figure.2).

The proposed Graphical interface manual physical sketch is seemed to be a very simple and non attractive graphical interface but it is not the final design, the user of the application is supposed to design the required graphical user interface itself.

### 3.2 Personal Assistant

The proposed physical sketch of Personal Assistant as shown in Figure 2, is designed keeping Agent based design requirements in mind. PA is proposed to be an agent based software application having autonomous behavior for working in a particular domain by communicating, moving and collaborating with users and other software applications in an automated environment to obtain desired results. The main job of the Personal Assistant in I-SOAS Graphical Interface is to communicate with user; take user's strutted and instructed requests, read them and forward them for father processing back to the user with obtained results.

## 4. INTERNAL WORK FLOW

Figure 3 shows the flow of the software. The interaction starts with the user input. At first the user inputted request is analyzed and then it's classified into two major categories .i.e., Build graphical user interface and Process Input.

If the user requested input is about to change the user out-look of graphical interface then the control will go to the Build graphical user interface category and if the user requested input is about to extract some information by processing and modeling data then the control will go to the process input step.

During Build graphical user interface operations, at first the requested user input is again analyzed and categorized into two further categories .i.e., Update graphical user interface and Build New graphical user interface. If the user's request is about to build a completely new graphical outlook then it will move to the step Build New graphical user interface but if the user's request is about to make some minor or major changes in already running outlook then it will move to the Update GUI. Final interface will then become the final output from I-SOAS Graphical Interface.

During Process Input operations user's requested input forwarded to the Processing and Modelling Component and final resultant output is obtained via Presentation Process Layer. This final output will then become the final resultant information for the user.

## 5. I-SOAS SYSTEM SEQUENCE

Figure 4 shows the Sequence design of the Graphical Interface System. There are four main components .i.e., Interface, Input Classification, Build or Modify graphical user interface and, Process Presentation Layer. The job of interface is to take input request from user and forward it to the input classification component which will verify, validate and classify the inputted user request. If user requested to make or update existing graphical interface then the input classification component will forward the user re-quest to Build of Modify graphical user interface component which will modify or build graphical user interface and will update the final changes on interface.

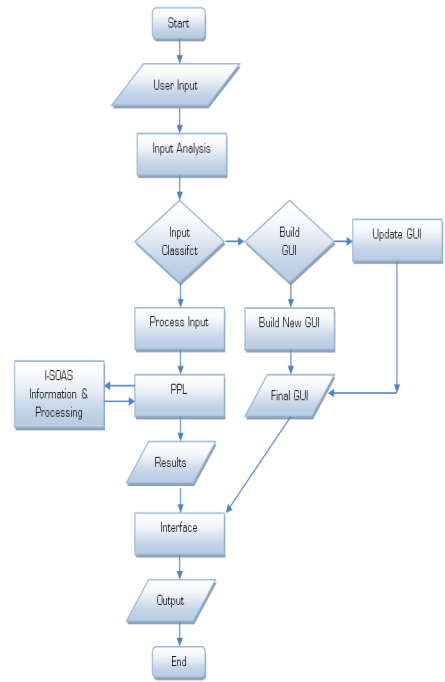

**Figure.3.** I-SOAS: Graphical Interface Internal Flow

Whereas if user requested to process and extract some information then input classification component will forward the request to Process Presentation Layer which will then forward the request to Information Processing and Modeling component for further processing and take back the resultant information to present on the graphical user interface.

## 6. TECHNOLOGIES INVOLVED

To implement I-SOAS Graphical Interface we are considering Java [4] as the main programming language and Rich Internet Applications [5] technologies .i.e., FLEX (Adobe) [6], AJAX [7] OpenLaszlo [8], Silverlight [9] .

We have chosen Rich Internet Applications because it provides small Installation footprint, reduces overhead for updating and distributing the application, provides automatic and transparent updates to new versions, provides usage of application from any computer with an internet connection, gives consistency to user regardless of what operating system the client and less prone to viral infection than running an actual executable. Moreover Rich Internet Applications provides features and functionalities for both the desktop and web application's human machine interface development.

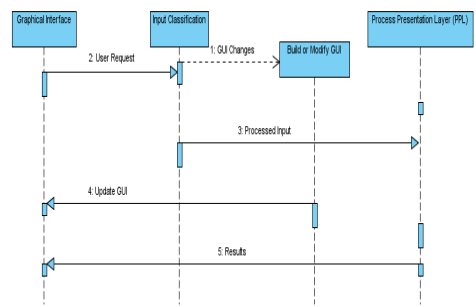

**Figure.4.** I-SOAS: Graphical Interface System Sequence Design

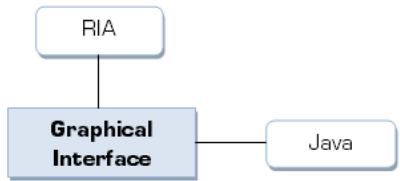

**Figure 5.** I- IUI-Technologies Involved

Traditional web applications centered all activities around client server architecture with a thin client where as Rich Internet Applications typically transfers the processing necessary for the user interface to the web client but keep the bulk of the data (i.e., maintaining the state of the program) back on the application server. Furthermore Rich Internet Applications is rich in user end functionality development options, more responsive than standard Web browsers, balanced frees server resources, asynchronous and efficient in reducing the network traffic.

## 7. I-SOAS PROTOTYPE

Following the designed implementation designs and meeting design requirements, we have started the implementation of I-SOAS.

As shown in the Figure 6, at the moment I-SOAS is available in prototype form, capable of running as a stable application, Taking input from user in the form of text format, analyzing user's natural language based instructions, providing flexibility in the organization and presentation of available options in the graphical machine interface and creating dynamic database to store information.

Moreover this implemented prototype version of I-SOAS Graphical Interface is flexible and capable of providing standard graphical interface, flexible graphical user interface (user can redesign and reconfigure the interface itself to accommodate specific needs by Mouse Click and Drag Drop options).

## 8. CONCLUSION

In this research paper we have briefly described Product Data Management and its some major existing challenges, then continuing the presentation of research briefly described the conceptual and implementable architecture of our own proposed solution based approach I-SOAS targeting PDM challenges.

In this research paper we have also described the design implementation of an intelligent, flexible and user friendly human machine interface .i.e., I-SOAS Graphical Interface (a component of I-SOAS) for the Product Data Management based applications human machine interface development. Moreover describing the detailed information about I-SOAS Graphical Interface we have presented information about the theme, design requirements, designed designs and technologies involved in the development of I-SOAS Graphical Interface.

As this research paper is about an ongoing in process research project, right now we are developing I-SOAS Graphical Interface by implementing above discussed proposed and designed designs.

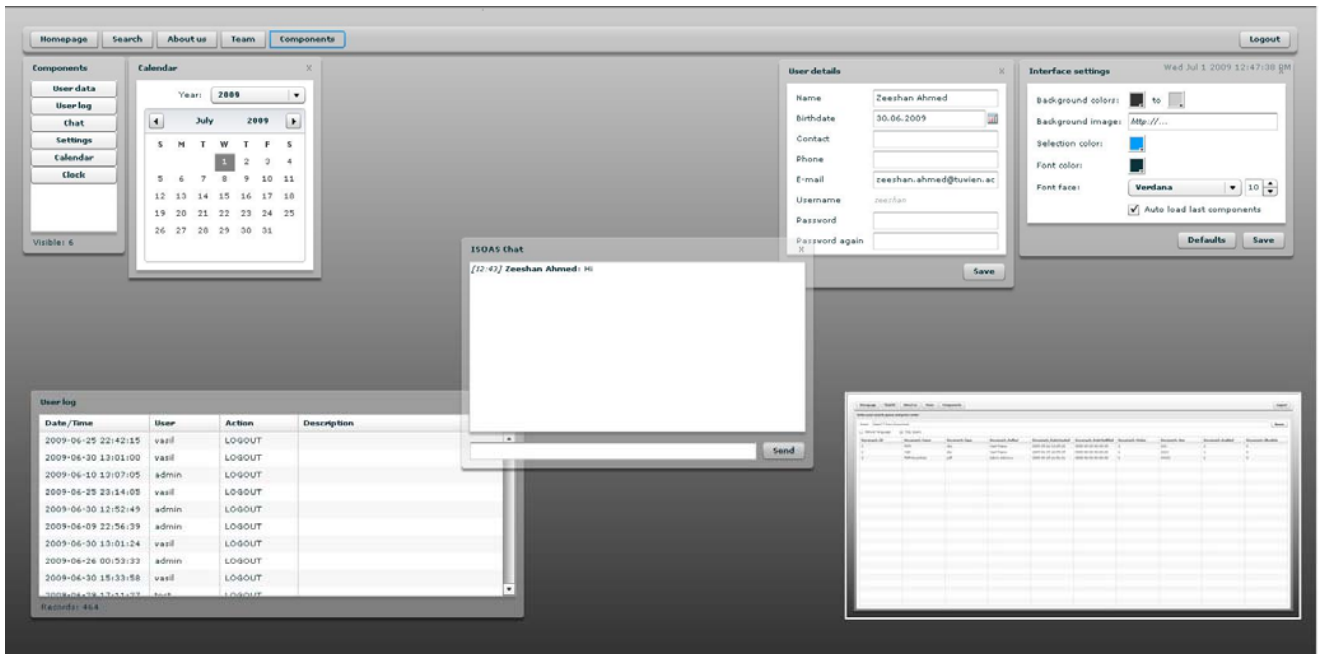

**Figure 6.** I-SOAS: Prototype